\documentclass[twocolumn,showpacs,amsmath,amssymb,prb]{revtex4}
\usepackage{graphicx,amstext}
\usepackage{dcolumn}
\usepackage{bm}
\usepackage[dvipdfm]{hyperref}
\newcommand{\vs}{{\it vs.\@}}

\newcommand{\al}{{\it et al.\@}}

\newcommand{\sr}{Sr$_{14-x}$Ca$_{x}$Cu$_{24}$O$_{41}$}
\newcommand{\bq}{\begin{equation}}
\newcommand{\eq}{\end{equation}}

\begin{document}

\title{Anisotropic Charge Modulation in Ladder Planes of Sr$_{14-x}$Ca$_{x}$Cu$_{24}$O$_{41}$ }

\author{T.~Vuleti\'{c}}
 \email{tvuletic@ifs.hr}
\author{T.~Ivek}
\author{B.~Korin-Hamzi\'{c}}
\author{S.~Tomi\'{c}}
 \homepage{http://www.ifs.hr/real_science}
\affiliation{Institut za fiziku, P.O.~Box 304, HR-10001 Zagreb, Croatia}
\author{B.~Gorshunov}
 \altaffiliation {Permanent address: General Physics Institute, Russian Academy
of Sciences, Moscow, Russia.}
\author{P.~Haas}
\author{M.~Dressel}
\affiliation{1.\@ Physikalisches Institut, Universit\"{a}t Stuttgart, D-70550
Stuttgart, Germany}
\author{J.~Akimitsu}
\author{T.~Sasaki}
\author{T.~Nagata}
\affiliation{Dept.~of Physics, Aoyama-Gakuin University, Kanagawa, Japan}

\date{\today}

\begin{abstract} 
The charge response of the ladders in \sr~ is characterized by dc resistivity, low
frequency dielectric and optical spectroscopy in all three crystallographic directions. The collective
charge-density wave screened mode is observed in the direction of the rungs for $x$$=$0, 3 and 6,
in addition to the mode along the legs. For $x$$=$8 and 9, the charge-density-wave response along
the rungs fully vanishes, while the one along the legs persists. The transport perpendicular to the planes is always dominated by hopping.
\end{abstract}

\pacs{{74.72.Jt}, {71.45.Lr}, {71.27.+a}, {78.70.Gq}}
\maketitle

Physics of doped Mott-Hubbard insulators challenges conventional theories of 
metals and insulators \cite{Imada84}. The effect of strong Coulomb interactions produces a 
rich variety of exotic ordering phenomena, which have been the focus of intense 
scientific activity in recent years. The spin-chain and ladder self-doped 
compound \sr~  has attracted much attention since it is the first 
superconducting copper oxide with a non-square lattice \cite{Nagata98}. Theoretically, in 
doped two-leg Cu-O ladders, superconductivity (SC) is tightly associated with 
the spin gap and in competition with charge-density wave (CDW) \cite{Dagotto99}. While both the 
spin gap  and CDW were established in the ladders of \sr~ \cite{Katano99,Gorshunov02,Blumberg02}, 
the relevance of these objects to electronic properties and superconductivity is 
still subject of intensive discussion. Recently, it was shown, on the basis of 
dielectric response data, that substitution of Sr$^{2+}$ by Ca$^{2+}$ gradually suppresses 
the insulating CDW phase, which eventually vanishes for $x$$>$9, Ref.\onlinecite{Vuletic03}. In contrast to 
these results, dynamical Raman response observed above RT for $x$$=$0 
was assigned to CDW fluctuations and found to persist in the metallic phase of 
$x$$=$12, a system which becomes SC under pressure \cite{Gozar03}. 

Of particular interest is to learn more about the nature of CDW order in the 
spin ladders, which present a nice experimental system of strongly interacting 
electrons. Although the ground state for $0\leq x \leq 9$ reveals a number of well-known 
fingerprints of the conventional CDW, such as the pinned phason \cite{Kitano01} and the 
broad dispersion at radio-frequencies due to screening of the CDW by free 
carriers~ \cite{Gorshunov02,Blumberg02,Vuletic03}, its origin certainly is more complicated, since the system does 
not undergo a {\em metal}-to-insulator but an {\em insulator}-to-insulator transition. The 
role of Ca-substitution is another open issue. Suppression of CDW phase was 
ascribed \cite{Vuletic03} to worsened nesting conditions \cite{Gruner88} implying that the system becomes 
more 2D already at ambient pressure for large x.  On the other hand, it was suggested 
that at ambient pressure (for all x) the charge dynamics is essentially 
one-dimensional (1D) \cite{Nagata98,Motoyama97} and that only the application of pressure induces 
the dimensional crossover from one to two \cite{Nagata98}.  

In this Communication, we address these important questions concerning the 
charge-ordered ground state in the ladders of \sr. Our results give 
evidence that the CDW is two-dimensional with an anisotropic dispersion: the 
long-range charge order develops only in ladder planes, leading to a screened 
collective response along the both legs and rungs of the ladders.

The dc resistivity of \sr~($x$$=$0, 3, 6, 8, 9 and 11.5) was investigated in the 
temperature range 2~K$< T <$700~K. The complex conductivity was measured in the 
frequency range 0.01~Hz$< \nu <$10~MHz, using several set-ups. At frequencies 
$\nu$$=$6 - 10000~cm$^{-1}$ the complex dielectric function was obtained by a 
Kramers-Kronig analysis of the infrared reflectivity and by measurements of the 
complex transmission coefficient at the lowest frequencies 6 - 20 cm$^{-1}$. All 
experiments were conducted on high-quality single crystals along the three 
crystallographic axes: $c$ (along the legs), $a$ (along the rungs) and $b$ 
(perpendicular to the ladder planes).

\begin{figure} 
\centering\includegraphics[clip,scale=0.53]{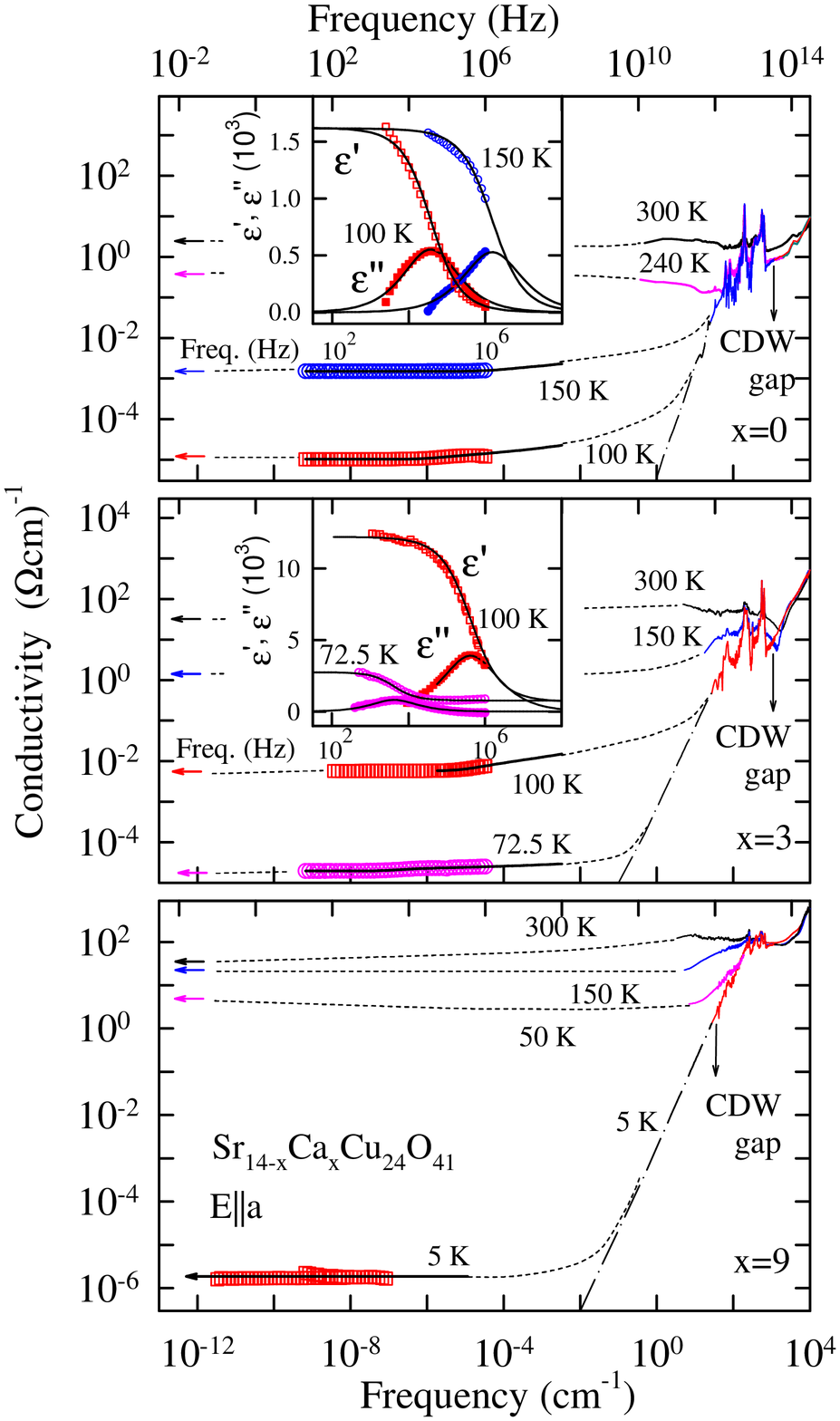} 
\caption{ (Color online)
Broad-band spectra of conductivity and complex dielectric function (real 
$\varepsilon$' and imaginary $\varepsilon$'' parts) of \sr~along the $a$-axis 
for Ca contents $x$$=$0, 3 and 9 at a few selected 
temperatures. Strong $T$-dependent relaxation-like dispersion of $\varepsilon$' 
and $\varepsilon$'' (insets, the full lines are from fits to the generalized 
Debye expression, see Text), seen also as smooth increase in the conductivity 
spectra, is a fingerprint of the screened CDW collective response. This response 
is observed at all $T<T_c$ for $x$$=$0 and 3, but not for $x$$=$9. Decrease of the 
infrared conductivity at low $T$ corresponds to the opening of an energy gap. At 
the lowest $T$ only the lowest-frequency phonon tail is seen and represented with 
the dash-dot $\nu^2$ line. The arrows denote the dc conductivity. Dotted lines 
are guides to the eye.}
\label{Fig1}
\end{figure}

\begin{figure}
\centering\includegraphics[clip,scale=0.46]{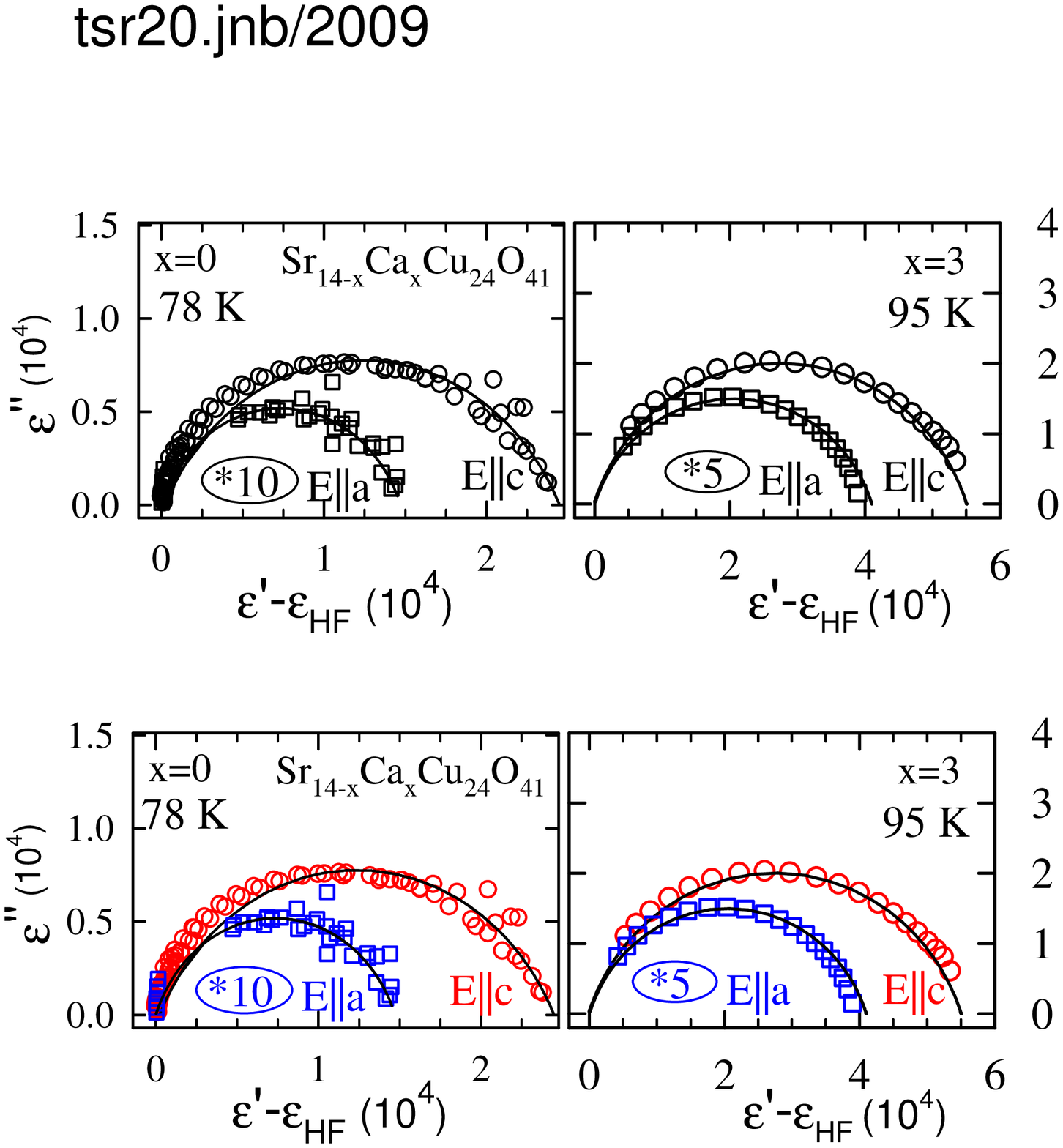}
\caption{
Representative Cole-Cole plots of the dielectric dispersion, connected to the 
anisotropic collective CDW response of \sr~for $x$$=$0 (left panel) and $x$$=$3 (right panel) with the $E||c$ (along 
legs) and $E||a$ (along rungs). Note that plots for the response along $a$-axis are 
blown-up 10 ($x$$=$0) and 5 ($x$$=$3) times.  Full lines are from fits to the generalized Debye 
expression. The intersections of the arcs with 
$\varepsilon$'$-\varepsilon_{HF}$ axes indicate the values of 
$\Delta\varepsilon$.}
\label{Fig2}
\end{figure}

Fig.~\ref{Fig1} shows the conductivity spectra in a broad frequency range parallel to the 
rungs, $E||a$. For $x$$=$0, 3, and 6 a strong $T$-dependent relaxation of the 
dielectric function  $\varepsilon(\omega)$$=$$\varepsilon$' + $i \varepsilon$'' 
is found (see insets). Fits by the generalized Debye expression  
$\varepsilon(\omega)$-$\varepsilon_{HF}$$=$$\Delta\varepsilon/[1+(i\omega\tau_0)
^{1-\alpha}]$ yield the main parameters of this relaxation: the dielectric 
strength $\Delta\varepsilon$$=$$\varepsilon_0$-$\varepsilon_{HF}$$\approx$$10^3$ 
($\varepsilon_0$ is static and $\varepsilon_{HF}$ is high-frequency dielectric 
constant), the symmetric broadening of the relaxation-time distribution given 
by $1-\alpha$$\approx$0.8, and the mean relaxation time $\tau_0$, which closely 
follows a thermally activated behavior similar to that of the dc conductivity. 
The dielectric response sets in below the phase transition temperatures 
$T_c$=210~K ($x$=0), 140~K ($x$=3), and 55~K ($x$=6) in the same manner as found 
parallel to the legs of the ladder (see Fig.2 in Ref.\onlinecite{Vuletic03}), only the dielectric 
strength $\Delta\varepsilon$  along the $a$ axis is much smaller than that for 
$E||c$ (Fig.~\ref{Fig2}). This analogy suggests that the same mechanism, {\em i.e.}~the 
screening of a CDW phason, is responsible for the ac properties in both 
directions of \sr, parallel to the legs and to the rungs of the ladders. For 
higher Ca content $x$$=$8 (not shown) and 9 the analogy breaks down since no 
such dispersion is detected for $E||a$ down to 4 K. In the third direction 
($E||b$), we find no signature of a CDW-related dielectric response at any Ca 
content.

\begin{figure} 
\centering\includegraphics[clip,scale=0.45]{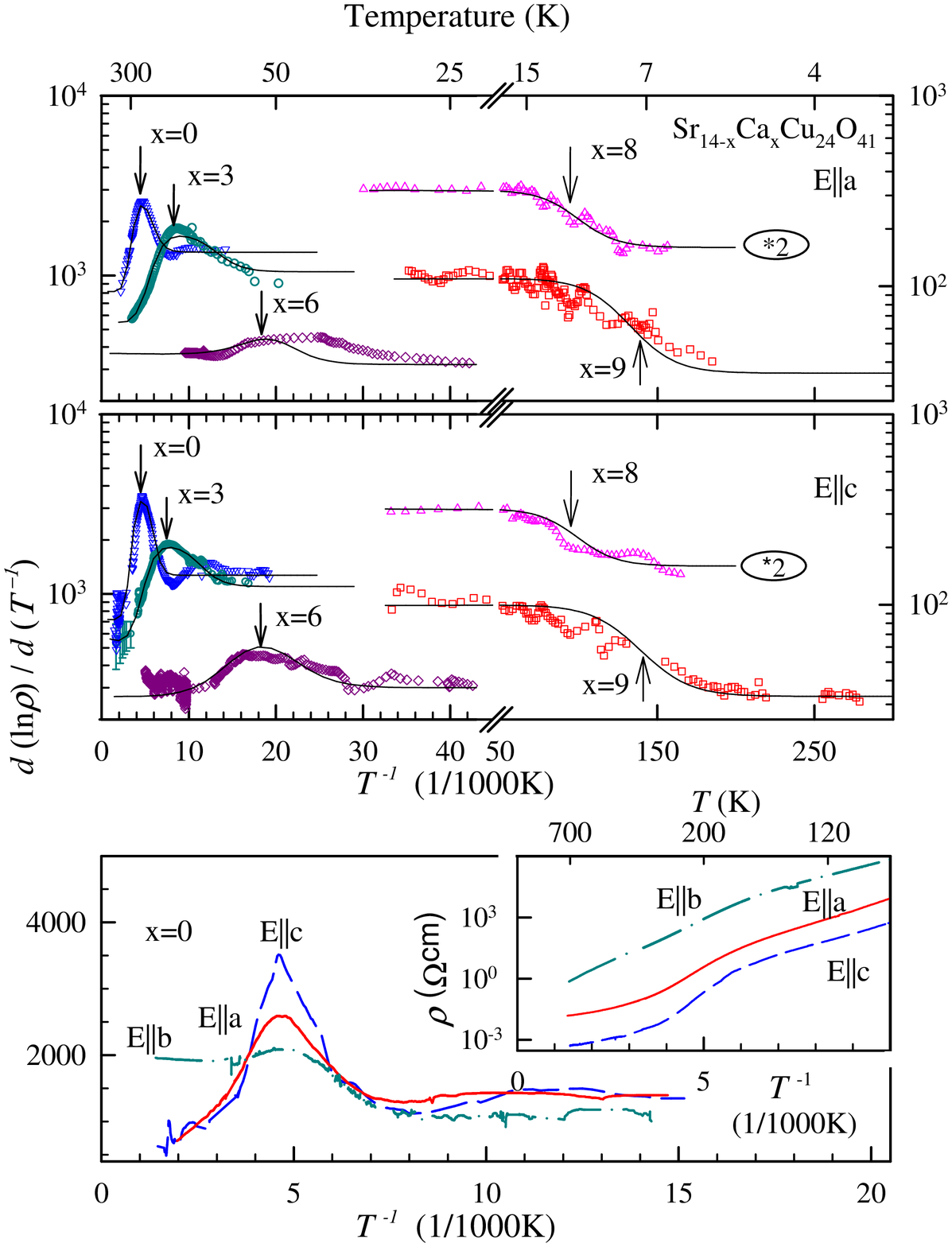} 
\caption{ 
(Color online) Upper and middle 
panels: dc resistivity logarithmic derivatives of \sr~for different Ca contents 
$x$ along the $c$- and $a$-axes;  $x$=8 is multiplied by 2, for 
clarity. Full lines are guides for the eye in the transition range, while above 
and below they are based on the fits to $\ln \rho$ \vs~$T^{-1}$. The arrows 
indicate the CDW phase transition temperature $T_c$. Lower panel: dc resistivity 
(Inset) and corresponding logarithmic derivatives of Sr$_{14}$Cu$_{24}$O$_{41}$ 
along $c$ (dashed), $a$ (solid) and $b$ (dash-dot line)  axes. The 
peak in $d (\ln \rho)/d(T^{-1})$ as a signature of the phase transition is 
observed in the ($c$, $a$) ladder plane, but not perpendicular to them (along 
$b$-axis).} 
\label{Fig3} 
\end{figure}

The energy gap associated with the CDW formation is also seen in the infrared 
spectra for $E||a$, where the conductivity at the lowest frequencies decreases 
upon cooling. The estimated gap values for different $x$ correspond well to 
those determined from the activated dc resistivity (Fig.~\ref{Fig3} and Table~\ref{TablART}); the 
values are also close to those found for $E||c$. It is worth of mentioning that 
Ruzicka \al \cite{Ruzicka99} have already suggested from their optical spectra the existence 
of the metal-to-insulator phase transition both along the $c$ and $a$ axes. The 
phonon bands become less pronounced for higher Ca contents $x$ due to 
screening by free carriers. At low $T$ only a  $\nu^2$ contribution (Fig.~\ref{Fig1}, dash-dot 
line) of the low-energy phonon wing is observed.

\begin{table}[!h]
\begin{tabular}{lrrrrr}
\hline
\hline
&$x$$=$0&$x$$=$3&$x$$=$6&$x$$=$8&$x$$=$9\\
\hline
$\Delta_{\mathrm {CDW}}$[meV]&130$\pm$5&110$\pm$5&30$\pm$4& 8$\pm$1& 3$\pm$0.5\\
$\Delta_{\mathrm {H.T.}}$~[meV]&90$\pm$25& 80$\pm$20&30$\pm$6&16$\pm$2&10$\pm$2\\
\hline
\hline
\end{tabular}
\caption{
CDW gaps, $\Delta_{\mathrm {CDW}}$, are from dc and ac measurements of \sr. 
Activation energies, $\Delta_{\mathrm {H.T.}}$, in the high-$T$ 
insulating phase, are from dc measurements. 
$\Delta_{\mathrm {CDW}}$ and $\Delta_{\mathrm {H.T.}}$  are obtained with 
both the $E||c$ and $E||a$.} 
\label{TablART} 
\end{table}

We now compare the dc transport properties along all three 
directions. The upper two panels of Fig.~\ref{Fig3} reveal that the phase transition 
temperatures $T_c$, when measured along the $c$-axis and $a$-axis, are equal in value 
and have the same dependence on Ca content $x$. The same holds for the energy 
gaps in the insulating phase above $T_c$ and in the CDW ground state, which both 
are isotropic and show the same dependence on $x$. The energy gaps 
become larger when going from high temperatures into the CDW phase, indicating 
that an additional gap opens for $x \lesssim 6$ (Table~\ref{TablART}). While in a standard CDW, a 
transition from a metallic state to an insulating state is observed due to the 
opening of an energy gap \cite{Gruner88}, in the present case the transport in the high-$T$ 
phase is already non-metallic. We explain this by electron-electron interactions 
within the ladder plane, leading towards a Mott insulator. We remind that the 
band filling in the ladders is close to 1/2 and the on-site Coulomb repulsion is 
$U$=3-4~eV, so that $U/4t$$>$1. CDW order in such a system of strongly interacting 
charges might fall between the two well-defined limits: the CDW 
order of itinerant charges and the charge order  of localized charges. The 
transition broadens substantially with increasing $x$, as reflected by an increase 
of the transition width and a decrease of the peak height of $d (\ln \rho)/d(T^{-1})$ (Fig.~\ref{Fig3}). 
The broadening might be attributed to disorder introduced by 
Ca-substitution; a well known effect in quasi-1D compounds \cite{Tomic91}. Finally, for 
the pure compound ($x$$=$0) the dc resistivity along the third axis $E||b$ (lower 
panel of Fig.~\ref{Fig3}) shows that the activation energy is much larger at high $T$ (190 
meV) and becomes smaller with decreasing temperature. For $x$$\geq$3 (not shown) there 
is a single activation in the whole $T$-range. This simple activation process 
indicates that the charge transport perpendicular to the ladder planes 
happens via nearest-neighbor hopping, as expected between disordered chains. 
Since no peak of $d (\ln \rho)/d(T^{-1})$ is found in $b$-direction, the CDW does not 
develop a long range order in 3D. These findings are in accord with our optical data, as well as data 
by Ruzicka \al~\cite{Ruzicka99}, which indicate $T$-independent insulating 
behavior along $b$-axis, distinct from the charge dynamics in the ladder planes.

Our observations have several implications. Generally, in a 1D metal one expects 
the development of a CDW only along the chain axis. Hence the existence of the 
loss peak as the signature of the screened phason relaxation in the 
perpendicular direction (along the rungs) is surprising 
\cite{Littlewood87}. However, in a 2D system of coupled chains, CDW develops 
according to the ordering vector $Q$. An ac electric field applied either 
parallel or perpendicular to the chains couples to the CDW developed along 
$Q^{-1}$ resulting in an anisotropic dispersion \cite{footnote2}. Indeed,  we  
find that the radio-frequency loss peak centered at $\tau_0^{-1}$ decays 
Arrhenius-like for both directions $E||c$ and $E||a$; these are well established 
features to characterize the CDW phason screened response. The dielectric 
strength of the loss-peak found along the $a$-axis is about 10 times smaller 
than the one observed along the $c$-axis, which corresponds to the 
single-particle conductivity anisotropy. Based on this anisotropic dispersion in 
the radio-frequency range, we may extrapolate from the standard phason 
dispersion in 1D, which connects the screened loss peak centered at  $\tau_0^{-1}$ 
to the unscreened pinned mode at $\Omega_0$, \cite{Vuletic03,Littlewood87}, to the 
2D case. Assuming that CDW effective mass anisotropy is given by the band mass 
anisotropy \cite{Gruner88}, we get an estimate of the pinning frequency of the 
CDW along the rungs:   $\Omega_0^2 (a) = \Omega_0^2 (c) \times [\sigma_{dc}(c)/ 
\sigma_{dc} (a)] \times [m_{band}(c)/ m_{band} (a)]$. Taking $m_{band}(c)/ 
m_{band} (a) \approx  0.1$, Ref.\onlinecite{Arai97},  $\Omega_0 (c) = 
1.5$-$1.8$~cm$^{-1}$, Ref.\onlinecite{Kitano01} and  $\sigma_{dc}(c)/ \sigma_{dc} (a) 
\approx 10$-30, we  find  $\Omega_0 (a) \approx 1.5$-$3.5$~cm$^{-1}$.  Indeed, a 
thorough analysis of the relevant data in Ref.\onlinecite{Kitano01}  identifies the pinned mode along rungs at 
$\Omega_0 (a) = 1.5$~cm$^{-1}$.

Our findings indicate that a CDW develops in the ($c$,$a$)-ladders plane with a 
2D long-range order. Further, we address the robustness of the CDW inside the 
phase diagram of \sr. Ca-substitution quickly suppresses the length scale at 
which the 2D CDW in the ladder planes is developed, as broadening of the CDW 
phase transition indicates. For $x$$=$8 the absence of the peak in dc 
resistivity logarithmic derivative suggests that the long-range order in planes 
is destroyed; subsequently the CDW is able to respond only along the legs of the 
ladders before disappearing completely from the phase diagram for $x$$>$9. The 
picture in which CDW fluctuations and gap of 185 meV persist for all x 
(including metallic $x$$=$12) at $T$$>$300~K, Ref.\onlinecite{Gozar03}, is difficult to 
reconcile with our data. This picture also meets difficulties because the 
changes of the unit cell parameters $a$ and $c$ (at RT), induced by Ca-substitution, are small and 
comparable in size \cite{Pachot99}.   More importantly, we find that the 
conductivity anisotropy  almost does not vary with Ca content at high-$T$; it is $T$-independent for 
$0$$\leq$$x$$\leq$$8$, and becomes enhanced  at low $T$ for $x$$=$9 and 
11.5, \cite{Nagata98,Motoyama97,Osafune99}. No increase in dimensionality indicates that standard nesting 
arguments cannot explain the suppression of the CDW by Ca-substitution. 
Therefore, we propose an alternative scenario based on the low doped Mott 
insulator nature of the high-$T$ phase from which CDW in the ladders originates. 
According to this scenario, CDW phase is suppressed by Ca-substitution by the 
deviation from half-filling (which might be induced by even slight increase of 
the hole transfer into the ladders), as well as by an increase in intraladder 
overlap integrals, when $U/4t$ decreases. Changes of the inter-site Coulomb 
repulsion, due to an increased coupling between ladders and chains 
\cite{Pachot99}, might also influence the stability of CDW phase. The similar 
rate by which charge order, associated with the 2D antiferromagnetic dimer 
pattern, in the chains is supressed, is striking \cite{Kataev01}. Same arguments 
can be applied to explain gradual suppression of the high-$T$ insulating phase. 
However, CDW phase is clearly less robust and the influence of disorder 
introduced by Ca-substitution at Sr sites plays an important role. Tsuchiizu 
\al~have recently derived the model for two-leg extended Hubbard ladder with 
both on-site and inter-site Coulomb repulsion \cite{Tsuchiizu04}. They show that 
decreasing the latter, together with increasing the doping, destabilizes CDW and 
$p$-density wave, which coexist in the phase diagram, in favor of $d$-SC 
state.

The charge-ordered phase in the ladders of \sr~belongs to the class 
of broken sym\-me\-try pat\-terns, pre\-dicted theo\-reti\-cally for strongly correlated 
electronic systems, including charge and spin density waves, both of the site 
and bond order, and of $d$-sym\-me\-try \cite{Tsuchiizu04,Wu03}. However, there is no theoretical prediction about collective 
excitations in these phases (except for CDW) and how they should respond 
to applied dc and ac fields. Wu \al~\cite{Wu03} showed that charge-ordered phases in 
two-leg ladders at low doping levels can develop only quasi-long-range order due 
to degeneracies appearing in the systems away from half-filling. Indeed, the CDW 
transition in the ladders of \sr~is ten times broader (even in the 
Ca-free compound, $x$$=$0) than expected \cite{Tomic91}, indicating that probably a true 
long-range order is never reached.  

In conclusion, we demonstrated that within the ladder planes of 
\sr, the charge undergoes a two-dimensional ordering, whose length 
scale is quickly reduced by Ca-substitution due to an increased disorder. 
Similar to a CDW, the collective excitations of this charge order possess an 
anisotropic phason-like dispersion, which we detect as broad screened relaxation 
modes along both the rungs and legs of the ladders. We propose that the 
charge-ordered phase vanishes at high Ca levels ($x>9$) due to an increased 
deviation from half-filling and an increase in intraladder overlap integrals, 
when $U/4t$ decreases. At this point, angle-resolved photoemission studies of 
momentum-resolved gaps and self-energies as a function of temperature and Ca-
substitution could help in understanding the mechanism responsible for producing 
this charge-ordered state. Also, further experiments have to clarify whether 
such a dispersion is a unique feature of the charge order in ladders or whether 
it is common to quasi-2D systems with charge order.

We thank A.~Bjeli\v{s}, P.~Littlewood and P.~Monceau for useful 
discussions and G.~Untereiner for the samples preparation. This work was 
supported by the Croatian Ministry of Science and the Deutsche 
Forschungsgemeinschaft.


\begin{thebibliography}{10}
\bibitem{footnote2}
The phason screened dielectric relaxation was only observed for ac electric 
field applied along the best conductivity axis of the quasi-1D compounds. In our 
best knowledge there was no attempt until now to measure it along the less 
conductivity axes.
\end{thebibliography}
\end{document}